\documentclass[aps,prd,preprintnumbers,nofootinbib,preprint]{revtex4}
\usepackage{bm}
\usepackage{latexsym}
\usepackage{dcolumn}
\usepackage{amsmath,amsfonts,amssymb}
\usepackage{graphicx}
\usepackage{color}
\usepackage{amsthm}
\usepackage{epstopdf}
\usepackage{hyperref}

\begin{document}
\title{   Short Distance Physics of the Inflationary de Sitter Universe }

\author{Ahmed Farag Ali$^{1,2}$}\email[]{afali@fsu.edu; ahmed.ali@fsc.bu.edu.eg}
\author{Mir Faizal$^3$} \email[]{f2mir@uwaterloo.ca}
\author{Mohammed M. Khalil$^{4}$} \email[]{moh.m.khalil@gmail.com}
\affiliation{$^1$Department of Physics, Florida State University, Tallahassee, FL 32306, USA}
\affiliation{$^2$Department of Physics, Faculty of Science, Benha University, Benha 13518, Egypt}
\affiliation{$^3$Department of Physics and Astronomy, University of Waterloo, Waterloo, Ontario, N2L 3G1, Canada}
\affiliation{$^4$Department of Electrical Engineering,\\ Alexandria University,  Alexandria 12544, Egypt}

\begin{abstract}
   In this work, we investigate inflationary cosmology using  scalar field theory deformed by the generalized uncertainty principle (GUP) containing a linear momentum term. Apart from being consistent with the existence of a minimum measurable length scale, this  GUP is  also consistent with doubly special relativity and hence with the existence of maximum measurable momentum. We use this deformed scalar field theory to analyze the tensor and scalar mode equations in a de Sitter background, and to calculate modifications to the tensor-to-scalar ratio. Finally, we compare our results for the tensor-to-scalar ratio with the Planck data to constrain the minimum length parameter in the GUP.
\end{abstract}
%\pacs{123xxx}

\maketitle

\section{Introduction}
Even though we do not have a full theory of quantum gravity, it is expected that various approaches to quantum gravity may be used to gain some phenomenological information about quantum gravitational processes. 
One common feature among most of these approaches,  including    string theory  and semi-classical black hole physics, is the existence of a minimal observable length
\cite{Amati:1988tn,Scardigli:1999jh,Garay:1994en,Kempf:1994su,Kempf:1996fz,Brau:1999uv,Maggiore:1993rv}.   The existence of a
minimal length leads to the modification of the Heisenberg uncertainty
principle to a Generalized Uncertainty Principle (GUP). Since the Heisenberg algebra is closely related to the Heisenberg uncertainty principle, the GUP deforms the position-momentum commutation relation by  an additional quadratic term in momentum 
\cite{Amati:1988tn,Scardigli:1999jh,Garay:1994en,Kempf:1994su,Kempf:1996fz,Brau:1999uv,Maggiore:1993rv},
\begin{equation}
\label{gupquadratic}
[x, p] = i(1+\beta p^2),
\end{equation}
where $\sqrt{\beta}$ parametrizes the minimum length.
Phenomenological aspects of GUP effects have been analyzed in several contexts such as the
self-complete character of gravity \cite{Isi:2013cxa}, the conjectured black hole productions at terascale \cite{Mureika:2011hg,Nicolini:2011nz} and the modifications of
neutrino oscillations \cite{Sprenger:2010dg}.

Recently, another interesting form of the GUP has been proposed in  \cite{Ali:2009zq,Das:2010zf,Ali:2011fa} that 
 contains  a linear term in momentum
\begin{equation}
\label{gup}
[x_i, p_j] = i \left[\delta_{ij}-\alpha\left( p\delta_{ij} +
\frac{p_i p_j}{p}\right) + \alpha^2 
\left(p^2 \delta_{ij}+3 p_{i} p_{j} \right)  \right]~
\end{equation}
where $\alpha$ parametrizes the minimum length. This form of GUP is 
consistent with doubly special relativity (DSR) and black hole physics. DSR \cite{AmelinoCamelia:2000mn, Magueijo:2001cr} is an extension to special relativity in which  there is  an observer independent maximum energy scale, usually the Planck energy, apart from the observer independent maximum velocity i.e., the velocity of light.  Since this GUP is consistent with DSR, this GUP implies  the existence of a maximum observable momentum, apart from the existence of a    minimum measurable length  \cite{Ali:2009zq,Das:2010zf,Ali:2011fa}. 

Based on some dimensional arguments, one would expect that the fundamental length scale would be the Planck scale, but this is not necessarily true. In string theory for example, the string length scale could be one or two orders of magnitude greater than the Planck length scale \cite{Kaplunovsky:1987rp}, or even more in the ADD model \cite{ArkaniHamed:1998rs,Antoniadis:1998ig}. This is why the minimum length scale from the GUP is parametrized by the free parameter $\alpha$.
Upper bounds on $\alpha$ have been calculated in \cite{Ali:2011fa,das2011planck},
and it was proposed that the GUP may introduce an intermediate length scale between
Planck scale and electro-weak scale. Recent proposals suggested that these bounds
could be measured using quantum optics techniques \cite{Pikovski:2011zk} or gravitational wave techniques \cite{Marin:2014wja,Marin:2013pga}. If this happens, then it can be   
considered a milestone in quantum gravity phenomenology. In a series of papers, various phenomenological implications of the new model of GUP were investigated
\cite{Majhi:2013koa,Amelino-Camelia:2013fxa,Majumder:afa,Nozari:2012gd,Ching:2012vq,Ali:2013qza,Ali:2011ap,Tawfik:2012he}.
A detailed review along the mentioned lines of minimal length theories and quantum gravity phenomenology can be found in \cite{Hossenfelder:2012jw,Sprenger:2012uc,AmelinoCamelia:2008qg}.

One possible area in which minimum length could yield observable signatures is inflation. In inflation, quantum fluctuations in the early universe gets stretched out to cosmological sizes, thus providing an explanations for structure formation from first principles \cite{mukhanov1981quantum,Hawking:1982cz,Bardeen:1983qw,guth1982fluctuations}. Inflationary models require a minimum of 60 e-folds of inflationary expansion to explain cosmological puzzles. This means that inflation expanded Planckian scales to astrophysical scales, and hence could act as a microscope for probing short distance scales. Thus, suggesting that Planck scales could have observable effects on the universe today.

However, there is the possibility that Planck scale effects gets diluted with the expansion of the universe. This is the case in black hole physics, where Hawking radiation does not carry imprints of trans-Planckian physics \cite{Brout:1995wp, Corley:1996ar}. In cosmology however, Brandenberger and Martin \cite{Brandenberger:2000wr} considered various models of trans-Planckian physics and found that some of those models yield some late time consequences while others do not. These models were based on modified dispersion relations.

Another approach is the model suggested in \cite{Kempf:2000ac, Kempf:2001fa} based on GUP \eqref{gupquadratic}, and was used to investigate the effect of minimum length on inflation.  Since the GUP is independent of an exact theory for quantum gravity, this approach   remains valid for many different theories for quantum gravity. 
The numerical  solution   for tensor perturbations  in this model has  been studied in \cite{Easther:2001fi, Easther:2001fz}, and it was observed that the effect on the power spectrum is of the order $\sqrt{\beta}H$. It has also been demonstrated in \cite{Ashoorioon:2004vm} that the presence of cutoff affects the boundary terms in addition to the bulk terms in the Lagrangian.
However, all this analysis was done using the GUP in Eq.\eqref{gupquadratic}. In this paper, we use the linear GUP \eqref{gup} for performing a similar analysis. We  calculate both the scalar and tensor perturbations, in addition to the tensor-to-scalar ratio for a field theory deformed by the linear GUP in de Sitter background. We compare our results for the tensor-to-scalar ratio with the Planck data and use it to rule out certain ranges for the parameter $\alpha$.

\section{Scalar and Tensor Fluctuations with Minimal Length}

In this section, we investigate how the action of a scalar field theory is deformed by the GUP. 
The action of scalar perturbations can be written as  \cite{Ashoorioon:2004vm,Ashoorioon:2004wd}
\begin{equation}
S_S^{(1)}=\frac{1}{2}\int d\eta d^3y z^2\left((\partial_\eta \phi)^2-\sum_{i=1}^{3}(\partial_{y^i}\phi)^2\right),
\end{equation}
where $\phi$ is the gauge-invariant intrinsic curvature,  $z=a^2\psi'_0/a'$, and $\psi'_0$ is the homogeneous part of the inflation field. An alternative way to write the action is
\begin{equation}
S_S^{(2)} =\frac{1}{2}\int d\eta d^3y\left((\partial_\eta u)^2-\delta^{ij} \partial_i u\partial_j u+\frac{z''}{z}u^2\right).
\end{equation} 
where $u=-z\phi$. Those two actions are equivalent in absence of minimal length up to a boundary term.

Similarly, the action for the tensor perturbations could take the form \cite{Ashoorioon:2004vm,Ashoorioon:2004wd}
\begin{equation}
S_T^{(1)}=\frac{M^2_P}{64\pi}\int d\eta d^3y a^2(\eta) \partial_\mu {h^i}_j \partial^\mu {h_i}^j,
\end{equation}
or
\begin{equation}
S_T^{(2)}=\frac{1}{2}\int d\eta d^3y \left(\partial_\eta {P_i}^j \partial^\eta {P^i}_j - \delta^{rs}\partial_r {P_i}^j\partial_s {P^i}_j+\frac{a''}{a}{P_i}^j{P^i}_j\right),
\end{equation}
where $h_{ij}$ is the transverse traceless part of tensor perturbations of the metric, and
\begin{equation}
{P^i}_j(y)=\sqrt{\frac{M_P^2}{32\pi}}a(\eta)h^i_j(y).
\end{equation}

In the presence of minimal length, the equations of motion were derived in \cite{Kempf:2000ac,Ashoorioon:2004vm} using the GUP \eqref{gupquadratic}. Here we review the derivation of the equation of motion from $S_S^{(1)}$ but using the linear GUP in Eq.\eqref{gup}.

Since the GUP implies minimal length, we transform to proper space  coordinates $x^i$ instead of the co-moving coordinates $y^i$
\begin{equation}
\label{actions1}
S=\int d\eta ~d^3x ~\frac{z^2}{2a^3}\left\{\left[\left(\partial_\eta + 
\frac{a'}{a}~\sum_{i=1}^3\partial_{x^i}x^i
 - \frac{3a'}{a}\right)\phi\right]^2 - a^2
\sum_{i=1}^3\left(
\partial_{x^i}\phi\right)^2\right\},
\end{equation}
where the prime denotes derivative with respect to $\eta$.
Now, by  defining
\begin{align}
(\phi_1,\phi_2) & :=  \int d^3x ~ \phi^*_1(x)\phi_2(x)\\
x^i\phi(x) & :=  x^i \phi(x)\\
p^i\phi(x) & :=  -i\partial_{x^i}\phi(x),
\end{align}
we can write the action in the form
\begin{equation}
\label{ab}
S=\int d\eta~\frac{z^2}{2a^3}\left\{\left(\phi,A^\dagger(\eta)
A(\eta)\phi\right)-a^2\left(\phi,p^2\phi\right)
\right\},
\end{equation}
where the operator $A(\eta)$ is defined as: 
\begin{equation}
\label{abac}
A= \left(\partial_\eta + i 
\frac{a'}{a}~\sum_{i=1}^3p^i x^i - \frac{3a'}{a}\right).
\end{equation}

In Eq.\eqref{ab}, the fields are time dependent 
abstract vectors in a Hilbert space representation 
of the commutation  relations
\begin{equation}
\label{qmccr}
[x^i,p^j]=i \delta^{ij}.
\end{equation}
We aim here to incorporate the GUP by modifying the underlying three-dimensional 
position-momentum commutation relations in Eq.\eqref{qmccr} with that in Eq. \eqref{gup}. 
A convenient Hilbert space representation 
of the new commutation relations
is on fields $\phi(\rho)$ over auxiliary variables $\rho^i$
\begin{equation}
x^i\phi(\rho) =  i\partial_{\rho^i}\phi(\rho),
\end{equation}
\begin{equation}
p^i\phi(\rho) = \frac{\rho^i}{1+\alpha \rho-2 \alpha^2{\rho}^2}\phi(\rho),
\end{equation}
with scalar product
\begin{equation}
(\phi_1,\phi_2) = \int_{{\rho}^2<\beta^{-1}}
d^3\rho~\phi_1^*(\rho)\phi_2(\rho).
\end{equation}

The action, as given in its abstract form, 
Eqs.\eqref{actions1} and \eqref{ab}, with the new  
commutation relation in Eq.\eqref{gup}
 underlying, can now be written in the $\rho$-representation as
\begin{equation}
S=\int d\eta\int_{{\rho}^2<\beta^{-1}}d^3\rho~\frac{z^2}{2a^3}\left\{
\left\vert \left(\partial_\eta -\frac{a'}{a}\frac{
\rho^i}{1+\alpha \rho-2\alpha^2 {\rho}^2}\partial_{\rho^i}
-\frac{3a'}{a}\right)\phi
\right\vert^2 -\frac{a^2 {\rho}^2\left\vert 
\phi\right\vert^2}{(1+\alpha \rho-2\alpha^2 {\rho}^2)^2} \right\}
\end{equation}

The $\rho$ modes are coupled, so we define a new coordinate system as follows 
\begin{equation}
\label{etank}
\tilde{\eta}=\eta, \qquad \tilde{k}^i= a \rho^i \exp(\alpha\rho-\alpha^2\rho^2),
\end{equation}
in which the $\tilde{k}$ modes decouple, because
\begin{equation}
\partial_{\bar\eta}= \partial_\eta- \frac{a^{\prime}}{a} \frac{\rho^i}{1+\alpha \rho- 2\alpha^2 \rho^2} \partial_{\rho^i}.
\end{equation}
With a few calculations, we obtain
\begin{equation}
\rho^2 d\rho= \tilde{k}^2 d\tilde{k} \frac{e^{-3(\alpha\rho-\alpha^2\rho^2)}}{a^3\left(1+\alpha\rho-2\alpha^2\rho^2\right)},
\end{equation}
and the action now reads 
\begin{equation}
S = \int d\eta \int_{\tilde{k}^2<a^2/e\beta}
 d^3\tilde{k}~ 
\frac{z^2}{2a^6}\nu
\left\{\left\vert\left(\partial_{\eta} - 
3\frac{a'}{a}\right)\phi_{\tilde{k}}(\eta)
\right\vert^2
- \mu
\vert\phi_{\tilde{k}}(\eta)
\vert^2\right\}
\label{nac1},
\end{equation}
with $\nu$ and $\mu$ defined by
\begin{equation}
\nu=\frac{e^{-3(\alpha\rho-\alpha^2\rho^2)}}{1+\alpha\rho-2\alpha^2\rho^2} \qquad
\mu=\frac{a^2\rho^2}{(1+\alpha\rho-2\alpha^2\rho^2)^2}.
\end{equation}

This would yield the modified equation of motion with new functions $\nu$ and $\mu$ as follows \cite{Ashoorioon:2004vm}
\begin{equation}
\label{eoms}
v''_{\tilde{k}}+\frac{\nu'}{\nu}v'_{\tilde{k}}+\left(\mu-\frac{z''}{z}-\frac{z'}{z}\frac{\nu'}{\nu}\right)v_{\tilde{k}}=0,
\end{equation}
where we defined $v_{\tilde{k}}=a^2\phi_{\tilde{k}}$.

It may be noted that the form of  the functions $\mu$ and $\nu$ is different from their form for the regular GUP \cite{Kempf:2000ac, Ashoorioon:2004vm}. Since these functions are important to determine the effect on GUP on inflation, we expect that the effect of the linear GUP on inflaton  will  be different from that of the quadratic GUP.  However, apart from the form of $\mu$ and $\nu$, the remaining calculations are very similar. So,  we could just use the expressions used for regular GUP  \cite{Ashoorioon:2004vm}  
\begin{align}
\label{eoms1}
v''_{\tilde{k}}+\frac{\nu'}{\nu}v'_{\tilde{k}}+\left(\mu-\frac{z''}{z}-\frac{z'}{z}\frac{\nu'}{\nu}\right)v_{\tilde{k}}&=0  &&\text{(derived from $S_S^{(1)}$)} \\
\label{eoms2}
v''_{\tilde{k}}+\frac{\nu'}{\nu}v'_{\tilde{k}}+\left(\mu-\frac{z''}{z}\right)v_{\tilde{k}}&=0  &&\text{(derived from $S_S^{(2)}$)} \\
\label{eomt1}
u''_{\tilde{k}}+\frac{\nu'}{\nu}u'_{\tilde{k}}+\left(\mu-\frac{a''}{a}-\frac{a'}{a}\frac{\nu'}{\nu}\right)u_{\tilde{k}}&=0  &&\text{(derived from $S_T^{(1)}$)} \\
\label{eomt2}
u''_{\tilde{k}}+\frac{\nu'}{\nu}u'_{\tilde{k}}+\left(\mu-\frac{a''}{a}\right)u_{\tilde{k}}&=0 && \text{(derived from $S_T^{(2)}$)}
\end{align}

In this paper, we use Eq. \eqref{eoms2} for scalar perturbations,  since $S_S^{(2)}$ is more commonly used in the literature because of its similarity with a massive free scalar field in Minkowski space-time \cite{Ashoorioon:2004wd}. For the tensor perturbations,  we use Eq. \eqref{eomt1} as was used in \cite{Easther:2001fi,Ashoorioon:2004wd}.  We consider the case of de Sitter space in which $a=-1/H\eta$, $z'/z=a'/a$, and $z''/z=a''/a$.

\section{Tensor Perturbations}

In this section, we analyze tensor perturbations deformed by linear GUP. 
The equation of motion for tensor perturbations can be written as  Eq. \eqref{eomt1}, 
\begin{equation}
\label{eomtensor}
u''_{\tilde{k}}+\frac{\nu'}{\nu}u'_{\tilde{k}}+\left(\mu-\frac{a''}{a}-\frac{a'}{a}\frac{\nu'}{\nu}\right)u_{\tilde{k}}=0
\end{equation}
We begin by expressing $\nu$ and $\mu$ in terms of $\tilde{k}$ which was defined in Eq.\eqref{etank}. To do that, we introduce the function $V$, which we define as the inverse of $xe^{-x-x^2}$ such that
\begin{equation}
V(xe^{-x-x^2})=x.
\end{equation}
From Eq. \eqref{etank}, this allows us to write $\rho$ in terms of $\tilde{k}$ as
\begin{equation}
\rho=-\frac{1}{\alpha}V(w),
\end{equation}
where we defined $w=-\alpha\tilde{k}/a$. Thus,  $\nu$ and $\mu$ could be written  as
\begin{equation}
\nu=\frac{e^{3(V(w)+V^2(w))}}{1-V(w)-2V^2(w)}, \qquad \mu=\frac{a^2V^2(w)}{\alpha^2\left(1-V(w)-2V^2(w)\right)^2},
\end{equation}
and
\begin{equation}
\frac{\nu'}{\nu}=\frac{a'}{a}\frac{V(w)\left(4+7V(w)-12V^2(w)-12V^3(w)\right)}{1-V(w)-2V^2(w)}.
\end{equation}

The equation of motion has a singularity at $w=-1$, because $V(-1)=-1$ and the denominator $1-V(-1)-2V^2(-1)=0$. This corresponds to conformal time $\eta=\eta_k$, where $\eta_k$ is when tensor perturbations with different co-moving momentum $k$ reach the cutoff $\rho=1/\alpha$. For de Sitter space $\eta=-1/Ha$, and hence $\eta_k=-1/\alpha \tilde{k}H$.

The $V$ function has a series expansion near $w=-1$ (see appendix)
\begin{equation}
V(w)=-1+p-\frac{1}{9}p^2+\frac{209}{648}p^3-...
\end{equation}
with $p=\sqrt{\frac{2}{3}(w+1)}$.
By repeating the analysis done for the regular GUP in \cite{Easther:2001fi}, we extract the most singular terms of the equation of motion to get the leading behavior of $u_{\tilde{k}}$
\begin{equation}
\label{eom2}
\ddot{u}_{\tilde{k}}+\frac{1}{2y}\dot{u}_{\tilde{k}}+\frac{1}{y}\left(\frac{1}{6\sigma^2}+\frac{1}{2}\right)u_{\tilde{k}}=0,
\end{equation}
where $\eta=\eta_k(1-y)$, $\sigma=\alpha H$, and the dot denotes derivative with respect to $y$. The parameter $\sigma$ is the ratio between the minimum length scale $\alpha$ and the Hubble length scale $H^{-1}$, and will be useful because in subsequent equations, $\alpha$ and $H$ are always multiplied by each other.

It may be noted that ignoring the $\dot{u}_{\tilde{k}}$ term, we obtain 
\begin{equation}
u''_k+\omega_k^2u_k=0,
\end{equation}
where $\omega_k^2=A/\eta_k^2y$, and
\begin{equation}
A=\frac{1}{6\sigma^2}+\frac{1}{2}.
\end{equation}
The solution is approximated by the WKB form
\begin{equation}
u_{\tilde{k}}(\eta)=\frac{1}{\sqrt{2\omega_k}}\exp(-i\int^\eta \omega_k(\eta')d\eta')
\end{equation}
This choice of vacuum is called the Bunch-Davies vacuum, and we get
\begin{equation}
\label{uk1}
u_k=\left(\frac{\eta_k^2y}{4A}\right)^{1/4}\exp(-2i\sqrt{Ay})
\end{equation}

Even though we cannot ignore the  $\dot{u}_{\tilde{k}}$ term, it is possible to use the solution obtained in Eq. \eqref{uk1} to obtain a deformed solution for the actual equation. Thus, we  
modify the prefactor of the exponential to obtain an exact solution for  Eq.\eqref{eom2}
\begin{equation}
\label{fy}
F(y)=\frac{\sqrt{A}}{2}\exp(-2i\sqrt{Ay}).
\end{equation}
The general solution for Eq. \eqref{eom2} can now be expressed as 
\begin{equation}
u_{\tilde{k}}(y)=C_+F(y)+C_-F^*(y),
\end{equation}
where the constants are constrained by the Wronskian condition \cite{Kempf:2000ac,Easther:2001fi}
\begin{equation}
u_{\tilde{k}}u'^{*}_{\tilde{k}}-u^*_{\tilde{k}}u'_{\tilde{k}}=i/\nu,
\end{equation}
leading to
\begin{equation}
|C_+|^2-|C_-|^2=-\frac{2\sqrt{6A}}{A^2}\eta_k y=\frac{2\sqrt{6A}}{A^2}\left(\frac{1}{\sigma k}+\eta\right). 
\end{equation}
It may be noted that this is $y$ dependent, but this is acceptable since we only need an approximate solution for Eq.\eqref{eomtensor} near the singularity to solve the equation numerically. Comparison with the Bunch-Davies solution suggests that $C_-=0$, which was confirmed in \cite{Easther:2001fi,Ashoorioon:2004wd}.

The solution \eqref{fy} is an approximate solution to Eq.\eqref{eomtensor}; we can get a more accurate solution  by extracting the subleading behavior of Eq.\eqref{eomtensor} using the method of dominant balance. Thus, using $u_{\tilde{k}}(y)=F(y)(1+\epsilon_1(y))$, we obtain  the equation
\begin{equation}
\ddot{\epsilon_1}+\frac{1}{2y}\dot{\epsilon_1}=i\frac{20}{27}\sqrt{\frac{3A}{2}}\frac{1}{y},
\end{equation}
which has the solution
\begin{equation}
\epsilon_1(y)=i\frac{20}{9}\sqrt{\frac{2A}{3}}y.
\end{equation}
Now we can mprove the solution by $u(y)=F(y)(1+\epsilon_1(y))(1+\epsilon_2(y))$, 
to obtain 
\begin{equation}
\ddot{\epsilon_2}+\frac{1}{2y}\dot{\epsilon_2}=-\frac{1}{\sqrt{y}}\left(\frac{40\sqrt{6}}{27}A-\frac{8}{27}\sqrt{\frac{3}{2}}A+\frac{400}{729}\sqrt{\frac{18}{2}}i\sqrt{A}+\frac{8}{9}\sqrt{\frac{3}{2}}\right). 
\end{equation}
The solution to this equation can be written as   
\begin{equation}
\epsilon_2(y)=-\frac{72\sqrt{6}}{243}(1+3A)y^{3/2}.
\end{equation}
Thus, the analytic solution near the singular point is given by
\begin{equation}
\label{u}
u_{\tilde{k}}=C_+F(y)(1+\epsilon_1(y))(1+\epsilon_2(y))+C_-F^*(y)(1+\epsilon_1(y))(1+\epsilon_2(y))
\end{equation} 

\begin{figure}[t]
	\centering
	\includegraphics[width=0.5\linewidth]{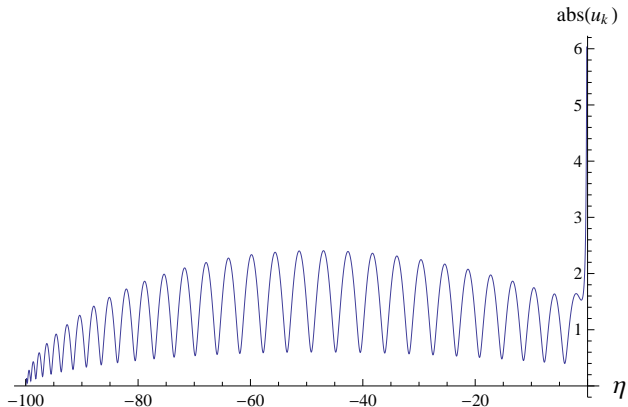}
	\caption{\label{fig:utabs}This figure shows how each mode evolves as a function of $\eta$ for $\sigma=0.01$. $|C_-|$ was taken to equal zero.}
\end{figure}

This solves the equation of motion near the singular point when the mode is well inside the horizon. To find the behavior of the tensor perturbations  when the mode is well outside the horizon, we solve the equation of motion numerically. This is done by evolving the analytic solution \eqref{u} when $\eta\to\eta_k$ numerically. In  Fig.\ref{fig:utabs}, we  plotted   the absolute value of $u_{\tilde{k}}$ for $\sigma=0.01$, and we see that it starts at a specific value for $\eta$, then oscillates and increases until it goes to infinity as $\eta$ goes to zero at present time.

The range of possible values for $\sigma=\alpha H$ is $\sigma<1$. 
This is because the lowest bound found on the GUP parameter $\alpha$ is $\alpha<10^8$ from the anomalous magnetic moment of the muon \cite{Das:2011tq}. In addition, during inflation, the universe expands exponentially such that $a\propto e^{Ht}$, and most inflationary models require a minimum of 60 e-folds of inflationary expansion, i.e. $Ht>60$, and that inflation ends after about $10^{-34}$s, or $10^{10}$ in Planck units. This means that approximately $H\sim 10^{-8}$. The bound on $\alpha$ and the value of $H$ during inflation suggest that $\sigma < 1$. Also,  $\sigma=1$ means the minimum length scale $\alpha$ equals the Hubble length scale $H^{-1}$, so its natural to investigate $\sigma<1$.

The condition for horizon crossing in terms of $\tilde{k}$ can be found from the relation between $\tilde{k}$ and $\rho$
\begin{equation}
\tilde{k}^i=a\rho^i e^{\alpha\rho-\alpha^2\rho^2}
\end{equation}
At the horizon,  we can express $\rho=H$.  Thus, in de Sitter space, we obtain
\begin{equation}
\eta_{\text{horizon}}=-\frac{1}{\tilde{k}}e^{\sigma-\sigma^2}.
\end{equation}
We are interested in the asymptotic values of $|u_{\tilde{k}}|$, when $\rho/H\to 0$, which we implemented in the numerical solution as  $\rho/H\to 0.01$
\begin{equation}
\eta_{\text{asymp}}=-\frac{0.01}{\tilde{k}}\exp(\sigma/100-\sigma^2/10000).
\end{equation}

\begin{figure}[t]
	\centering
	\begin{minipage}[b]{0.45\linewidth}
		\includegraphics[width=\linewidth]{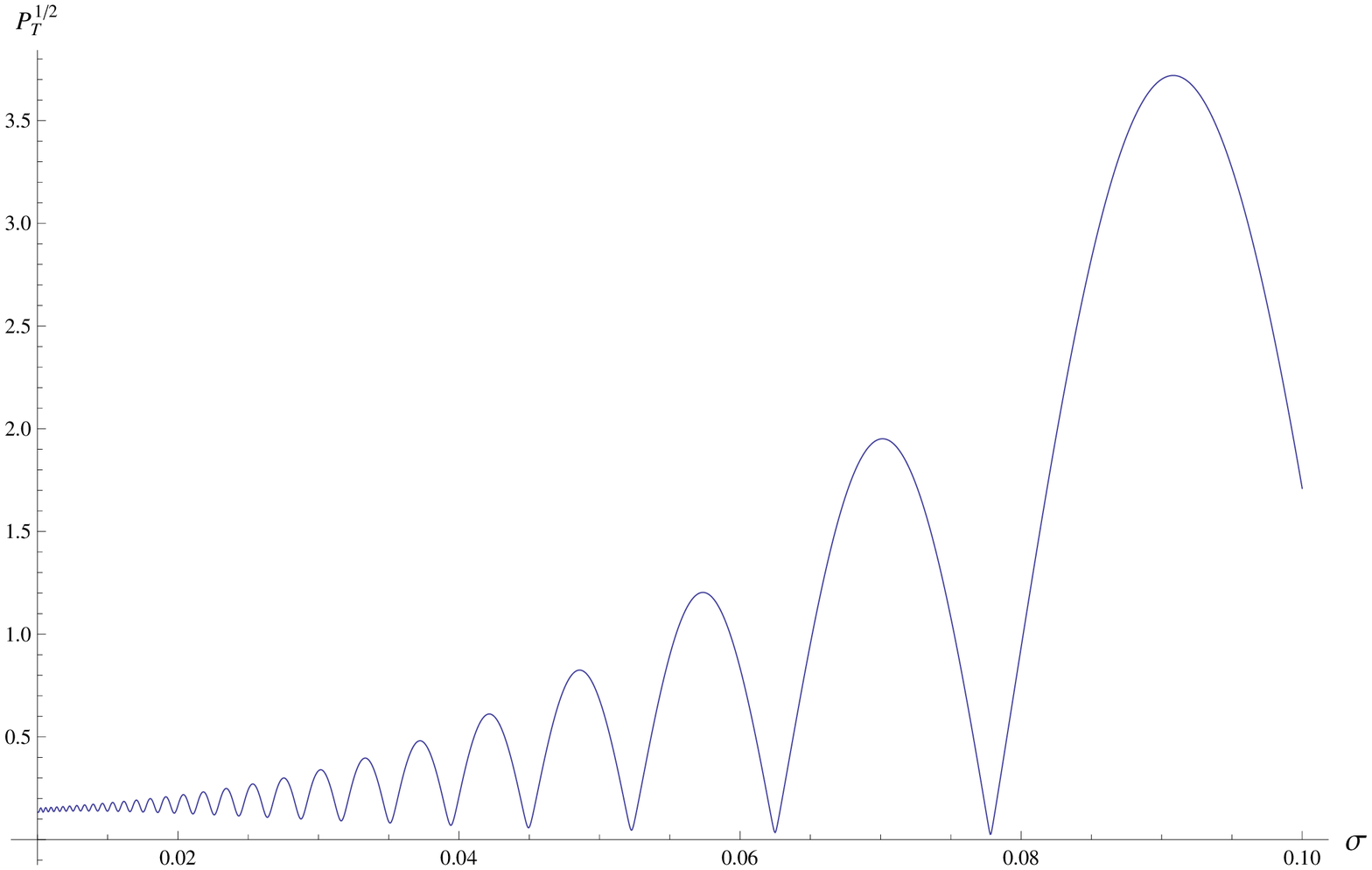}		
	\end{minipage}
	\quad
	\begin{minipage}[b]{0.45\linewidth}
		\includegraphics[width=\linewidth]{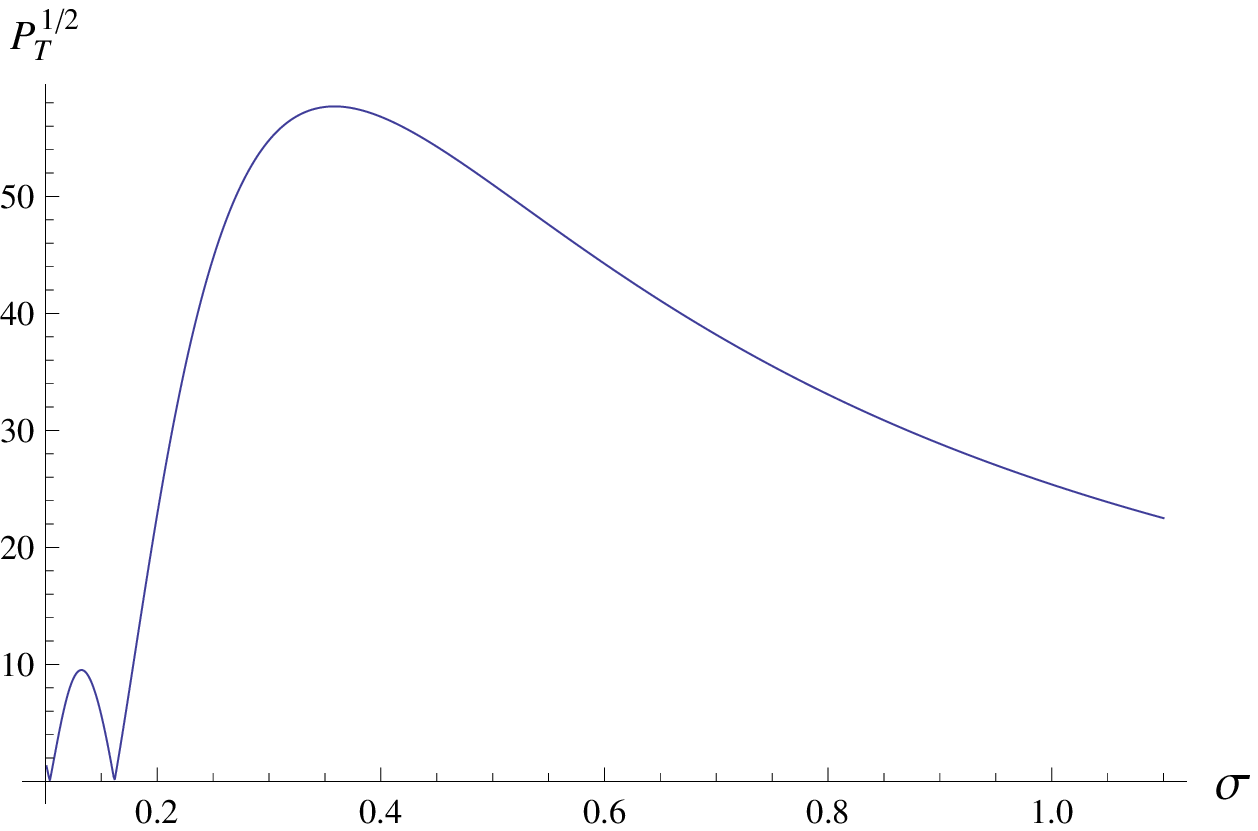}		
	\end{minipage}
	\caption{\label{fig:Pt} The dependence of the tensor power spectrum $P_T^{1/2}$ on $\sigma$ when $C_-=0$. The plot is independent of the value of $\tilde{k}$, and we obtain the standard result of $P_T^{1/2}=0.159155$ when $\alpha\to0$}
\end{figure}

The tensor power spectrum  in near de Sitter space can be expressed as \cite{Easther:2001fi}
\begin{equation}
P_T^{1/2}=\sqrt{\frac{\tilde{k}^3}{2\pi^2}}\left\lvert\frac{u_{\tilde{k}}}{a}\right\rvert. 
\end{equation}
This  is evaluated at the asymptotic value $\eta_{\text{asymp}}$. 
In Fig.\ref{fig:Pt},  we plotted the dependence of the tensor power spectrum on $\sigma$,  when $C_-=0$. The plot is independent of $\tilde{k}$, which was checked by using different values for $\tilde{k}$ and getting the same results. We obtain the standard result of $P_T^{1/2}=0.159155$ when $\alpha\to0$.

\section{Scalar Perturbations}
In this section, we analyze scalar perturbations deformed by linear GUP. 
The equation of motion for scalar perturbations \eqref{eoms2} can be written as 
\begin{equation}
\label{scalareom}
\ddot{v}_{\tilde{k}}+\frac{\nu'}{\nu}\dot{v}_{\tilde{k}}+\left(\mu-\frac{a''}{a}\right)v_{\tilde{k}}=0,
\end{equation}
where in de Sitter space $z''/z=a''/a$.

Following the same steps as for tensor perturbations, we extract the most singular terms as
\begin{equation}
\ddot{v}_{\tilde{k}}+\frac{1}{2y}\dot{v}_{\tilde{k}}+\frac{1}{6\sigma^2y}v_{\tilde{k}}=0,
\end{equation}
which has the exact solution
\begin{equation}
G(y)=\frac{\sqrt{B}}{2}e^{-2i\sqrt{By}},
\end{equation}
where we defined
\begin{equation}
B=\frac{1}{6\sigma^2}.
\end{equation}
Thus, the general solution around the singularity is
\begin{equation}
v_{\tilde{k}}=D_+ G(y) + D_- G^*(y),
\end{equation}
with the Wronskian constraint
\begin{equation}
|D_+|^2-|D_-|^2=-\frac{2\sqrt{6B}}{B^2}\eta_k y.
\end{equation}

To find an approximate solution to Eq.\eqref{scalareom}, we use the method of dominant balance by substituting $v_{\tilde{k}}=G(y)(1+\xi_1(y))(1+\xi_2(y))$ in Eq.\eqref{scalareom} to get the two equations
\begin{equation}
\ddot{\xi}_1+\frac{1}{2y}\dot{\xi}_1=i\frac{20}{27}\sqrt{\frac{3}{2}}\sqrt{B}\frac{1}{y},
\end{equation}
and
\begin{equation}
\ddot{\xi}_1+\frac{1}{2y}\dot{\xi}_1+\left(\frac{72}{27}\sqrt{\frac{3}{2}}B+i\frac{400}{243}\sqrt{B}\right)\frac{1}{\sqrt{y}}=0,
\end{equation}
which have the solutions
\begin{equation}
\xi_1(y)=i\frac{20}{9}\sqrt{\frac{2B}{3}}y,
\end{equation}
and
\begin{equation}
\xi_2(y)=-\frac{8}{729}\left(81\sqrt{6}B+100i\sqrt{B}\right)y^{3/2}.
\end{equation}
Thus, we obtain the following general solution 
\begin{equation}
v_{\tilde{k}}=D_+G(y)(1+\xi_1(y))(1+\xi_2(y))+D_-G^*(y)(1+\xi_1(y))(1+\xi_2(y)).
\end{equation}

\begin{figure}[t]
	\centering
	\includegraphics[width=0.5\linewidth]{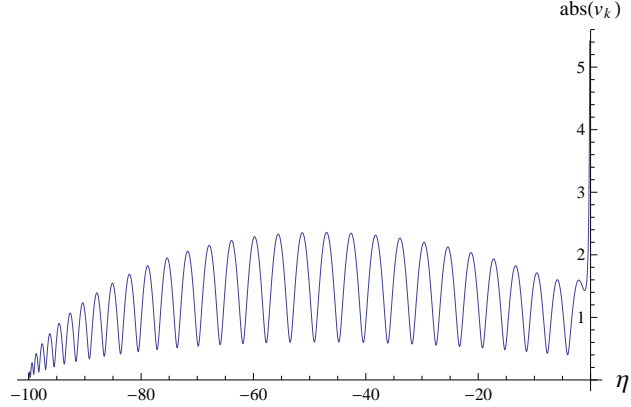}
	\caption{\label{fig:vsabs}This figure shows how each mode evolves as a function of $\eta$ for $H=0.1$ and $\alpha=0.1$. $|D_-|$ was taken to equal zero.}
\end{figure}

\begin{figure}[t]
	\centering
	\begin{minipage}[b]{0.45\linewidth}
		\includegraphics[width=\linewidth]{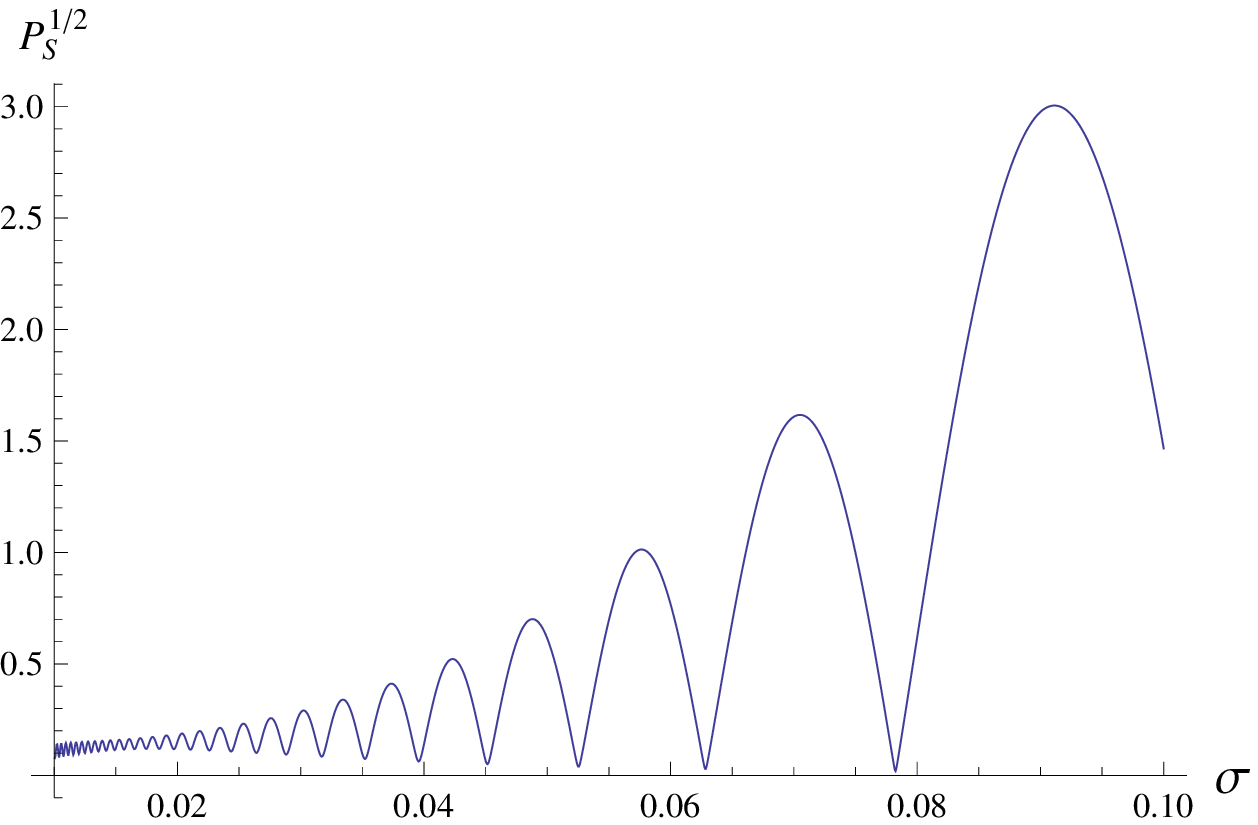}		
	\end{minipage}
	\quad
	\begin{minipage}[b]{0.45\linewidth}
		\includegraphics[width=\linewidth]{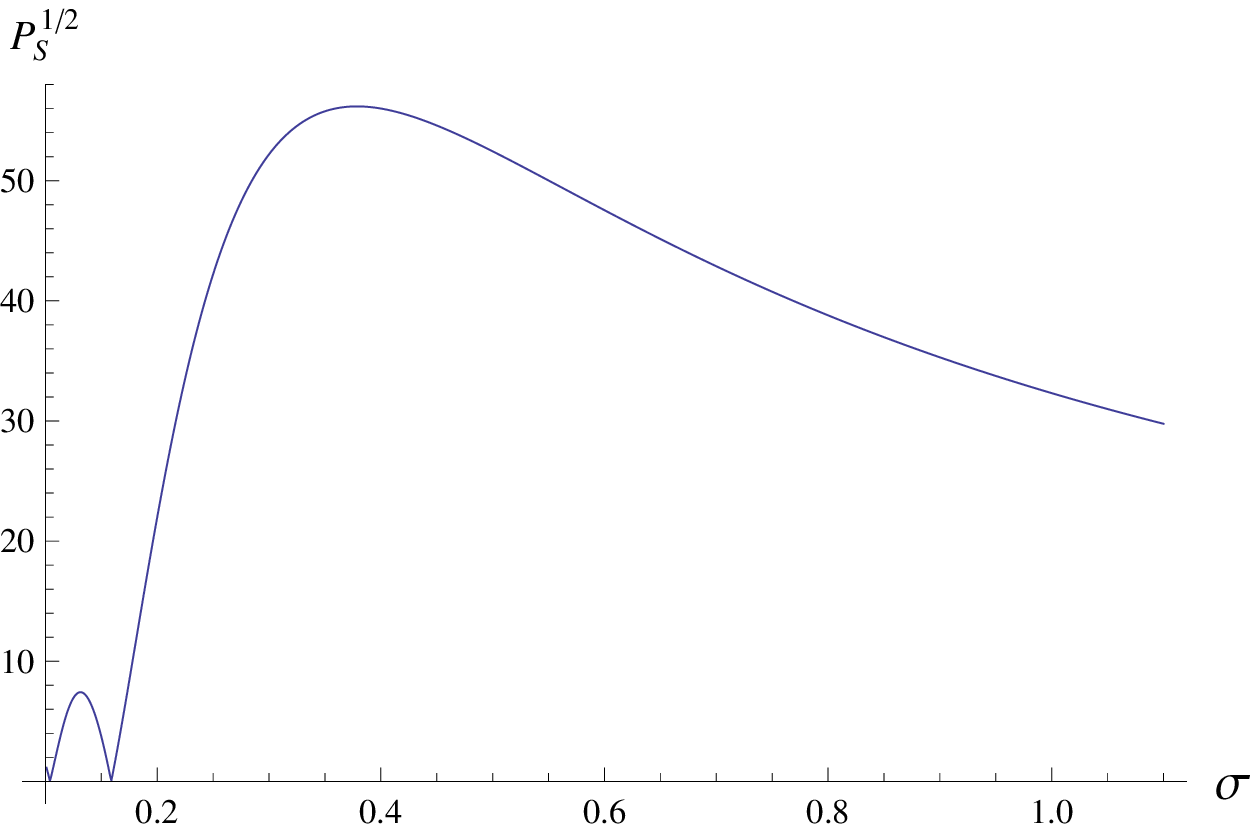}		
	\end{minipage}
	\caption{\label{fig:Ps}The dependence of the scalar power spectrum $P_S^{1/2}$ on $\alpha$ when $H=1$ and $C_-=0$. The power spectrum is independent of the value of $\tilde{k}$, and we obtain the standard result of $P_S^{1/2}=0.159155$ when $\alpha\to0$}
\end{figure}

We evolve this solution numerically to obtain the behavior of the scalar modes after crossing the horizon.
The scalar power spectrum is given by
\begin{equation}
P_S^{1/2}=\sqrt{\frac{\tilde{k}^3}{2\pi^2}}\left\lvert\frac{v_{\tilde{k}}}{z}\right\rvert
\end{equation}
In Fig.\ref{fig:vsabs}, we  plotted  $|v_{\tilde{k}}|$, and in Fig.\ref{fig:Ps}, we  plotted  the scalar power spectrum as a function of $\sigma$. Both plots are qualitatively similar to those of tensor perturbations but slightly shifted.

\section{Tensor-to-Scalar Ratio}
The tensor to scalar ratio is given by
\begin{equation}
r=\frac{A_T^2}{A_S^2},
\end{equation}
where $A_T$ and $A_S$ are the tensor and scalar amplitudes respectively, and are defined by \cite{Lidsey:1995np}
\begin{equation}
A_T(k)=\frac{1}{10}P_T^{1/2}, \qquad A_S(k)=\frac{2}{5}P_S^{1/2}.
\end{equation}

\begin{figure}[t]
	\centering
	\begin{minipage}[b]{0.45\linewidth}
		\includegraphics[width=\linewidth]{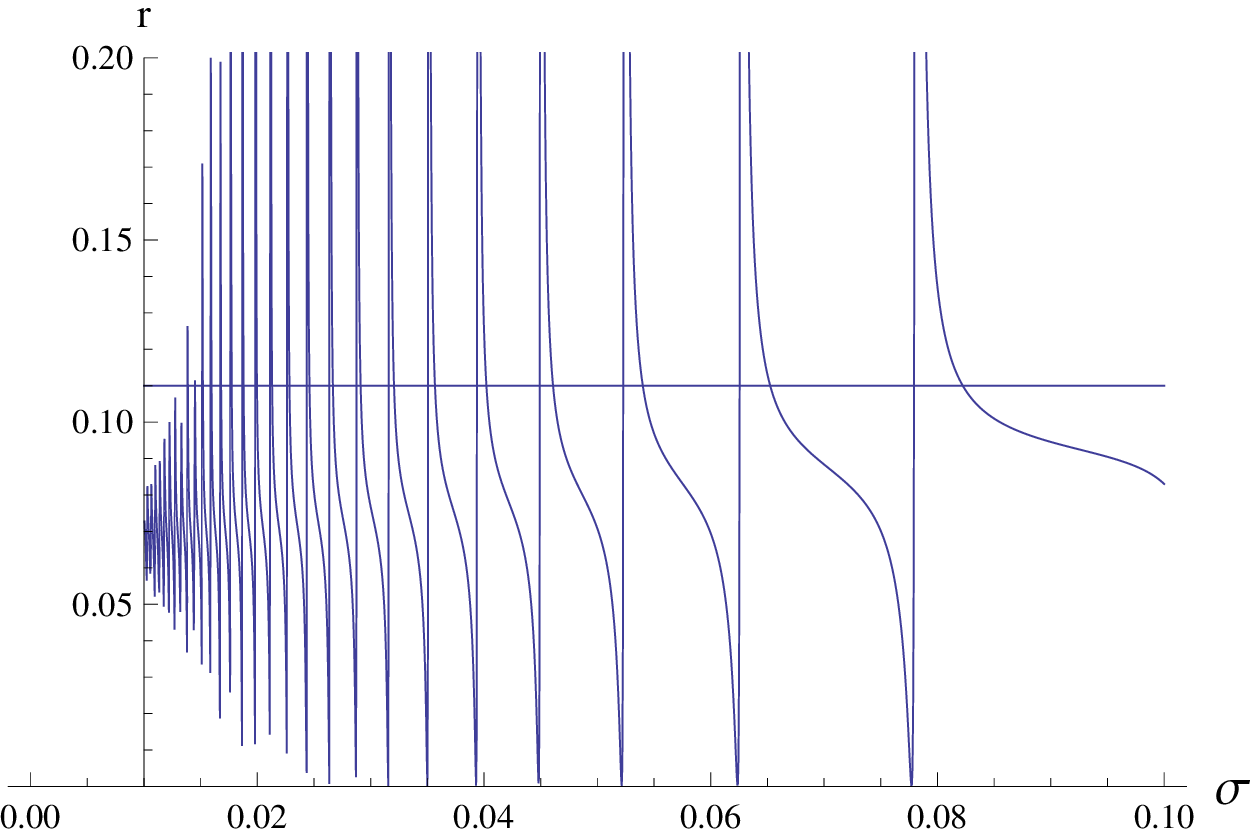}		
	\end{minipage}
	\quad
	\begin{minipage}[b]{0.45\linewidth}
		\includegraphics[width=\linewidth]{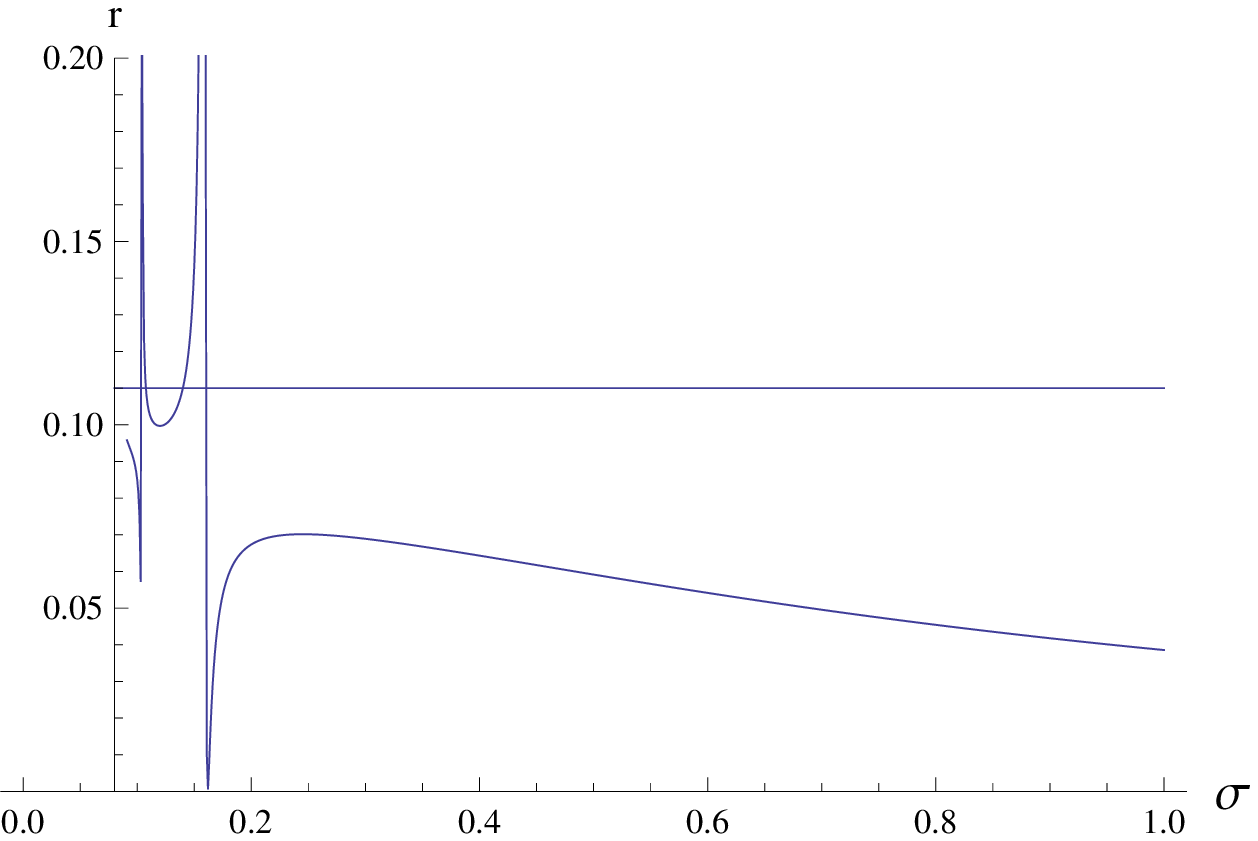}		
	\end{minipage}
	\caption{\label{fig:PtPs}Tensor to scalar ratio $r$ as a function of $\sigma$ for $|C_-|=|D_-|=0$.}
\end{figure}

In Fig.\ref{fig:PtPs},  we show the tensor to scalar ratio $r$ as a function of $\sigma$ for $|C_-|=|D_-|=0$. 
For small values of $\alpha$, $r$ oscillates rapidly about its standard value, and at certain values for $\alpha$ it gets much larger than the standard value, but $r$ decreases monotonically for $\sigma > 0.2$. We can see why these rapid oscillations occur by looking at Figs. \ref{fig:Pt} and \ref{fig:Ps} for $P_T^{1/2}$ and $P_S^{1/2}$ respectively. These two figures are similar, but slightly shifted from each other. So at the values of $\sigma$ when $P_T^{1/2}$ gets close to zero, the ratio $r$ also gets close to zero. But at the values of $\sigma$ when $P_S^{1/2}$ gets close to zero, the ratio $r$ increases rapidly.

The upper bound on $r$ from the Planck 2015 results \cite{Ade:2015lrj} is $r<0.11$. The value $r=0.11$ corresponds to the horizontal line in Fig.\ref{fig:PtPs}, and we see that it excludes some values for $\sigma$, and since $\sigma=\alpha H$, this constrains the possible values of the minimum length scale $\alpha$ once $H$ is determined in a specific inflationary model.

The tensor to scalar ratio for the GUP \eqref{gupquadratic} has  been investigated in \cite{Ashoorioon:2004wd}. It was demonstrated that the value of $r$ oscillates around the standard result, however the size of these oscillations tends to increase without a bound. So, the value of $r$ becomes  bigger than the Planck data bound on it for large values of the GUP parameter.

\section{Conclusions}
Inflationary models are quite sensitive to  short distance  physics, which motivated us to use a GUP modified framework for analyzing inflation. The GUP deformation we chose is consistent with doubly special relativity. This GUP introduces an additional linear term in the Heisenberg  algebra, and implies both a minimum length scale and a maximum momentum.  We investigated  the deformation of a scalar field theory in curved background by this GUP. We solved the tensor and scalar mode equations in de Sitter background, calculated the modified tensor-to-scalar ratio, and compared our results  with the Planck data.

It may be noted that it is possible that the Higgs field plays a role of the inflaton \cite{1w, w1}. Since the Higgs field is related to the  
electroweak sector of the standard model of particle physics, it is 
possible that inflation can occur in that sector of the standard model. It is possible to generalize the gauge principle for theories with a minimum length scale \cite{2w}, which  is done by first analyzing the full spacetime deformation caused by the generalized uncertainty principle, and then converting all the derivatives in the theory to gauge covariant derivatives. Such a generalization for the electroweak sector of standard model has also been performed in \cite{4w}. It will be interesting to use this theory to study the Higgs inflation with a minimum length scale. It may be noted that motivated by the linear GUP,  a non-local gauge theory has been constructed \cite{5w}. However, even though this theory is non-local, it can be effectively treated as a local theory in the framework of harmonic extension of functions. It would be interesting to generalize these results to electroweak sector and then analyze the Higgs inflation in this model.

\appendix*
\section{The function $V(x)$}
We introduced the function $V(x)$, which we define as the inverse of $xe^{-x-x^2}$, such that
\begin{equation}
V(xe^{-x-x^2})=x,
\end{equation}
or equivalently
\begin{equation}
V(x)e^{-V(x)-V(x)^2}=x.
\end{equation}

To find series expansion of the function $V(x)$ near any point, one could use the Lagrange inversion theorem. However, we need the expansion of the series around $x=-1$, and $V(x)$ is not analytic at this point since
\begin{equation}
\frac{dV}{dx}=\frac{V(x)}{x(1-V(x)-2V^2(x))}
\end{equation}

Following the method used in \cite{corless1996lambertw}, for the Lambert W function, we put $p=\sqrt{2(x+1)/3}$ in $Ve^{-V-V^2}=x$, and expand in powers of $1+V$
\begin{equation}
\frac{3}{2}p^2-1=-1+\frac{3}{2}(1+V)^2+\frac{1}{3}(1+V)^3-\frac{7}{8}(1+V)^4-\frac{3}{10}(1+V)^5+...
\end{equation}
Then, this series can be inverted to give
\begin{equation}
V(x)=-1+p-\frac{1}{9}p^2+\frac{209}{648}p^3-\frac{1537}{14580}p^4+...
\end{equation}

\begin{figure}[t]
	\centering
	\includegraphics[width=0.45\linewidth]{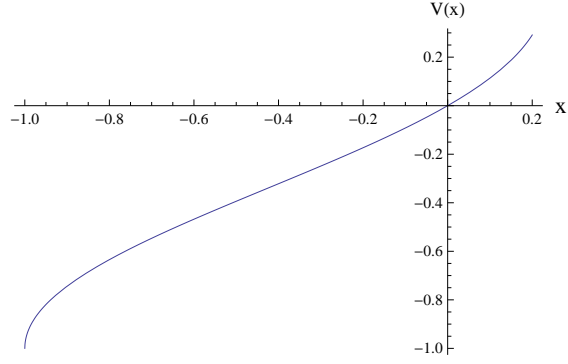}
	\caption{\label{fig:VPlot}A plot of $V(x).$}
\end{figure}

To evaluate $V(x)$ at any point for the numerical solution, we used Halley's method. This method is used to solve a nonlinear equation $f(x)=0$ by the sequence of iterations
\begin{equation}
x_{n+1}=x_n-\frac{2f(x_n)f'(x_n)}{2(f'(x_n))^2-f(x_n)f''(x_n)}.
\end{equation}
This method can be used to find successive approximations to $x=Ve^{-V-V^2}$ by
\begin{equation}
V_{n+1}=V_n-\frac{Ve^{-V-V^2}-x}{e^{-V-V^2}(1-V-2V^2) -\frac{(Ve^{-V-V^2}-x)(4V^3+4V^2-5V-2)}{2(1-V-2V^2)}}.
\end{equation}
In Fig.\ref{fig:VPlot}, we  plot   the $V$ function based on this numerical approximation.

\begin{acknowledgments}
	The research of AFA is supported by Benha University (www.bu.edu.eg). The authors would like to thank the anonymous referee for constructive comments and suggestions that significantly helped to improve this paper.
\end{acknowledgments}

\bibliography{gupinflation}

\begin{thebibliography}{56}
\expandafter\ifx\csname natexlab\endcsname\relax\def\natexlab#1{#1}\fi
\expandafter\ifx\csname bibnamefont\endcsname\relax
  \def\bibnamefont#1{#1}\fi
\expandafter\ifx\csname bibfnamefont\endcsname\relax
  \def\bibfnamefont#1{#1}\fi
\expandafter\ifx\csname citenamefont\endcsname\relax
  \def\citenamefont#1{#1}\fi
\expandafter\ifx\csname url\endcsname\relax
  \def\url#1{\texttt{#1}}\fi
\expandafter\ifx\csname urlprefix\endcsname\relax\def\urlprefix{URL }\fi
\providecommand{\bibinfo}[2]{#2}
\providecommand{\eprint}[2][]{\url{#2}}

\bibitem[{\citenamefont{Amati et~al.}(1989)\citenamefont{Amati, Ciafaloni, and
  Veneziano}}]{Amati:1988tn}
\bibinfo{author}{\bibfnamefont{D.}~\bibnamefont{Amati}},
  \bibinfo{author}{\bibfnamefont{M.}~\bibnamefont{Ciafaloni}},
  \bibnamefont{and}
  \bibinfo{author}{\bibfnamefont{G.}~\bibnamefont{Veneziano}},
  \bibinfo{journal}{Phys. Lett.} \textbf{\bibinfo{volume}{B216}},
  \bibinfo{pages}{41} (\bibinfo{year}{1989}).

\bibitem[{\citenamefont{Scardigli}(1999)}]{Scardigli:1999jh}
\bibinfo{author}{\bibfnamefont{F.}~\bibnamefont{Scardigli}},
  \bibinfo{journal}{Phys.Lett.} \textbf{\bibinfo{volume}{B452}},
  \bibinfo{pages}{39} (\bibinfo{year}{1999}), \eprint{hep-th/9904025}.

\bibitem[{\citenamefont{Garay}(1995)}]{Garay:1994en}
\bibinfo{author}{\bibfnamefont{L.~J.} \bibnamefont{Garay}},
  \bibinfo{journal}{Int.J.Mod.Phys.} \textbf{\bibinfo{volume}{A10}},
  \bibinfo{pages}{145} (\bibinfo{year}{1995}), \eprint{gr-qc/9403008}.

\bibitem[{\citenamefont{Kempf et~al.}(1995)\citenamefont{Kempf, Mangano, and
  Mann}}]{Kempf:1994su}
\bibinfo{author}{\bibfnamefont{A.}~\bibnamefont{Kempf}},
  \bibinfo{author}{\bibfnamefont{G.}~\bibnamefont{Mangano}}, \bibnamefont{and}
  \bibinfo{author}{\bibfnamefont{R.~B.} \bibnamefont{Mann}},
  \bibinfo{journal}{Phys.Rev.} \textbf{\bibinfo{volume}{D52}},
  \bibinfo{pages}{1108} (\bibinfo{year}{1995}), \eprint{hep-th/9412167}.

\bibitem[{\citenamefont{Kempf}(1997)}]{Kempf:1996fz}
\bibinfo{author}{\bibfnamefont{A.}~\bibnamefont{Kempf}},
  \bibinfo{journal}{J.Phys.} \textbf{\bibinfo{volume}{A30}},
  \bibinfo{pages}{2093} (\bibinfo{year}{1997}), \eprint{hep-th/9604045}.

\bibitem[{\citenamefont{Brau}(1999)}]{Brau:1999uv}
\bibinfo{author}{\bibfnamefont{F.}~\bibnamefont{Brau}},
  \bibinfo{journal}{J.Phys.} \textbf{\bibinfo{volume}{A32}},
  \bibinfo{pages}{7691} (\bibinfo{year}{1999}), \eprint{quant-ph/9905033}.

\bibitem[{\citenamefont{Maggiore}(1993)}]{Maggiore:1993rv}
\bibinfo{author}{\bibfnamefont{M.}~\bibnamefont{Maggiore}},
  \bibinfo{journal}{Phys.Lett.} \textbf{\bibinfo{volume}{B304}},
  \bibinfo{pages}{65} (\bibinfo{year}{1993}), \eprint{hep-th/9301067}.

\bibitem[{\citenamefont{Isi et~al.}(2013)\citenamefont{Isi, Mureika, and
  Nicolini}}]{Isi:2013cxa}
\bibinfo{author}{\bibfnamefont{M.}~\bibnamefont{Isi}},
  \bibinfo{author}{\bibfnamefont{J.}~\bibnamefont{Mureika}}, \bibnamefont{and}
  \bibinfo{author}{\bibfnamefont{P.}~\bibnamefont{Nicolini}},
  \bibinfo{journal}{JHEP} \textbf{\bibinfo{volume}{1311}}, \bibinfo{pages}{139}
  (\bibinfo{year}{2013}), \eprint{1310.8153}.

\bibitem[{\citenamefont{Mureika et~al.}(2012)\citenamefont{Mureika, Nicolini,
  and Spallucci}}]{Mureika:2011hg}
\bibinfo{author}{\bibfnamefont{J.}~\bibnamefont{Mureika}},
  \bibinfo{author}{\bibfnamefont{P.}~\bibnamefont{Nicolini}}, \bibnamefont{and}
  \bibinfo{author}{\bibfnamefont{E.}~\bibnamefont{Spallucci}},
  \bibinfo{journal}{Phys.Rev.} \textbf{\bibinfo{volume}{D85}},
  \bibinfo{pages}{106007} (\bibinfo{year}{2012}), \eprint{1111.5830}.

\bibitem[{\citenamefont{Nicolini and Winstanley}(2011)}]{Nicolini:2011nz}
\bibinfo{author}{\bibfnamefont{P.}~\bibnamefont{Nicolini}} \bibnamefont{and}
  \bibinfo{author}{\bibfnamefont{E.}~\bibnamefont{Winstanley}},
  \bibinfo{journal}{JHEP} \textbf{\bibinfo{volume}{1111}}, \bibinfo{pages}{075}
  (\bibinfo{year}{2011}), \eprint{1108.4419}.

\bibitem[{\citenamefont{Sprenger et~al.}(2011)\citenamefont{Sprenger, Bleicher,
  and Nicolini}}]{Sprenger:2010dg}
\bibinfo{author}{\bibfnamefont{M.}~\bibnamefont{Sprenger}},
  \bibinfo{author}{\bibfnamefont{M.}~\bibnamefont{Bleicher}}, \bibnamefont{and}
  \bibinfo{author}{\bibfnamefont{P.}~\bibnamefont{Nicolini}},
  \bibinfo{journal}{Class.Quant.Grav.} \textbf{\bibinfo{volume}{28}},
  \bibinfo{pages}{235019} (\bibinfo{year}{2011}), \eprint{1011.5225}.

\bibitem[{\citenamefont{Ali et~al.}(2009)\citenamefont{Ali, Das, and
  Vagenas}}]{Ali:2009zq}
\bibinfo{author}{\bibfnamefont{A.~F.} \bibnamefont{Ali}},
  \bibinfo{author}{\bibfnamefont{S.}~\bibnamefont{Das}}, \bibnamefont{and}
  \bibinfo{author}{\bibfnamefont{E.~C.} \bibnamefont{Vagenas}},
  \bibinfo{journal}{Phys.Lett.} \textbf{\bibinfo{volume}{B678}},
  \bibinfo{pages}{497} (\bibinfo{year}{2009}), \eprint{0906.5396}.

\bibitem[{\citenamefont{Das et~al.}(2010)\citenamefont{Das, Vagenas, and
  Ali}}]{Das:2010zf}
\bibinfo{author}{\bibfnamefont{S.}~\bibnamefont{Das}},
  \bibinfo{author}{\bibfnamefont{E.~C.} \bibnamefont{Vagenas}},
  \bibnamefont{and} \bibinfo{author}{\bibfnamefont{A.~F.} \bibnamefont{Ali}},
  \bibinfo{journal}{Phys.Lett.} \textbf{\bibinfo{volume}{B690}},
  \bibinfo{pages}{407} (\bibinfo{year}{2010}), \eprint{1005.3368}.

\bibitem[{\citenamefont{Ali et~al.}(2011)\citenamefont{Ali, Das, and
  Vagenas}}]{Ali:2011fa}
\bibinfo{author}{\bibfnamefont{A.~F.} \bibnamefont{Ali}},
  \bibinfo{author}{\bibfnamefont{S.}~\bibnamefont{Das}}, \bibnamefont{and}
  \bibinfo{author}{\bibfnamefont{E.~C.} \bibnamefont{Vagenas}},
  \bibinfo{journal}{Phys.Rev.} \textbf{\bibinfo{volume}{D84}},
  \bibinfo{pages}{044013} (\bibinfo{year}{2011}), \eprint{1107.3164}.

\bibitem[{\citenamefont{Amelino-Camelia}(2002)}]{AmelinoCamelia:2000mn}
\bibinfo{author}{\bibfnamefont{G.}~\bibnamefont{Amelino-Camelia}},
  \bibinfo{journal}{Int.J.Mod.Phys.} \textbf{\bibinfo{volume}{D11}},
  \bibinfo{pages}{35} (\bibinfo{year}{2002}), \eprint{gr-qc/0012051}.

\bibitem[{\citenamefont{Magueijo and Smolin}(2002)}]{Magueijo:2001cr}
\bibinfo{author}{\bibfnamefont{J.}~\bibnamefont{Magueijo}} \bibnamefont{and}
  \bibinfo{author}{\bibfnamefont{L.}~\bibnamefont{Smolin}},
  \bibinfo{journal}{Phys.Rev.Lett.} \textbf{\bibinfo{volume}{88}},
  \bibinfo{pages}{190403} (\bibinfo{year}{2002}), \eprint{hep-th/0112090}.

\bibitem[{\citenamefont{Kaplunovsky}(1988)}]{Kaplunovsky:1987rp}
\bibinfo{author}{\bibfnamefont{V.~S.} \bibnamefont{Kaplunovsky}},
  \bibinfo{journal}{Nucl. Phys.} \textbf{\bibinfo{volume}{B307}},
  \bibinfo{pages}{145} (\bibinfo{year}{1988}), \bibinfo{note}{[Erratum: Nucl.
  Phys.B382,436(1992)]}, \eprint{hep-th/9205068}.

\bibitem[{\citenamefont{Arkani-Hamed et~al.}(1998)\citenamefont{Arkani-Hamed,
  Dimopoulos, and Dvali}}]{ArkaniHamed:1998rs}
\bibinfo{author}{\bibfnamefont{N.}~\bibnamefont{Arkani-Hamed}},
  \bibinfo{author}{\bibfnamefont{S.}~\bibnamefont{Dimopoulos}},
  \bibnamefont{and} \bibinfo{author}{\bibfnamefont{G.~R.} \bibnamefont{Dvali}},
  \bibinfo{journal}{Phys. Lett.} \textbf{\bibinfo{volume}{B429}},
  \bibinfo{pages}{263} (\bibinfo{year}{1998}), \eprint{hep-ph/9803315}.

\bibitem[{\citenamefont{Antoniadis et~al.}(1998)\citenamefont{Antoniadis,
  Arkani-Hamed, Dimopoulos, and Dvali}}]{Antoniadis:1998ig}
\bibinfo{author}{\bibfnamefont{I.}~\bibnamefont{Antoniadis}},
  \bibinfo{author}{\bibfnamefont{N.}~\bibnamefont{Arkani-Hamed}},
  \bibinfo{author}{\bibfnamefont{S.}~\bibnamefont{Dimopoulos}},
  \bibnamefont{and} \bibinfo{author}{\bibfnamefont{G.~R.} \bibnamefont{Dvali}},
  \bibinfo{journal}{Phys. Lett.} \textbf{\bibinfo{volume}{B436}},
  \bibinfo{pages}{257} (\bibinfo{year}{1998}), \eprint{hep-ph/9804398}.

\bibitem[{\citenamefont{Das and Mann}(2011{\natexlab{a}})}]{das2011planck}
\bibinfo{author}{\bibfnamefont{S.}~\bibnamefont{Das}} \bibnamefont{and}
  \bibinfo{author}{\bibfnamefont{R.}~\bibnamefont{Mann}},
  \bibinfo{journal}{Physics Letters B} \textbf{\bibinfo{volume}{704}},
  \bibinfo{pages}{596} (\bibinfo{year}{2011}{\natexlab{a}}).

\bibitem[{\citenamefont{Pikovski et~al.}(2012)\citenamefont{Pikovski, Vanner,
  Aspelmeyer, Kim, and Brukner}}]{Pikovski:2011zk}
\bibinfo{author}{\bibfnamefont{I.}~\bibnamefont{Pikovski}},
  \bibinfo{author}{\bibfnamefont{M.~R.} \bibnamefont{Vanner}},
  \bibinfo{author}{\bibfnamefont{M.}~\bibnamefont{Aspelmeyer}},
  \bibinfo{author}{\bibfnamefont{M.}~\bibnamefont{Kim}}, \bibnamefont{and}
  \bibinfo{author}{\bibfnamefont{C.}~\bibnamefont{Brukner}},
  \bibinfo{journal}{Nature Phys.} \textbf{\bibinfo{volume}{8}},
  \bibinfo{pages}{393} (\bibinfo{year}{2012}), \eprint{1111.1979}.

\bibitem[{\citenamefont{Marin et~al.}(2014)\citenamefont{Marin, Marino,
  Bonaldi, Cerdonio, Conti et~al.}}]{Marin:2014wja}
\bibinfo{author}{\bibfnamefont{F.}~\bibnamefont{Marin}},
  \bibinfo{author}{\bibfnamefont{F.}~\bibnamefont{Marino}},
  \bibinfo{author}{\bibfnamefont{M.}~\bibnamefont{Bonaldi}},
  \bibinfo{author}{\bibfnamefont{M.}~\bibnamefont{Cerdonio}},
  \bibinfo{author}{\bibfnamefont{L.}~\bibnamefont{Conti}},
  \bibnamefont{et~al.}, \bibinfo{journal}{New J.Phys.}
  \textbf{\bibinfo{volume}{16}}, \bibinfo{pages}{085012}
  (\bibinfo{year}{2014}).

\bibitem[{\citenamefont{Marin et~al.}(2013)\citenamefont{Marin, Marino,
  Bonaldi, Cerdonio, Conti et~al.}}]{Marin:2013pga}
\bibinfo{author}{\bibfnamefont{F.}~\bibnamefont{Marin}},
  \bibinfo{author}{\bibfnamefont{F.}~\bibnamefont{Marino}},
  \bibinfo{author}{\bibfnamefont{M.}~\bibnamefont{Bonaldi}},
  \bibinfo{author}{\bibfnamefont{M.}~\bibnamefont{Cerdonio}},
  \bibinfo{author}{\bibfnamefont{L.}~\bibnamefont{Conti}},
  \bibnamefont{et~al.}, \bibinfo{journal}{Nature Phys.}
  \textbf{\bibinfo{volume}{9}}, \bibinfo{pages}{71} (\bibinfo{year}{2013}).

\bibitem[{\citenamefont{Majhi and Vagenas}(2013)}]{Majhi:2013koa}
\bibinfo{author}{\bibfnamefont{B.~R.} \bibnamefont{Majhi}} \bibnamefont{and}
  \bibinfo{author}{\bibfnamefont{E.~C.} \bibnamefont{Vagenas}},
  \bibinfo{journal}{Phys.Lett.} \textbf{\bibinfo{volume}{B725}},
  \bibinfo{pages}{477} (\bibinfo{year}{2013}), \eprint{1307.4195}.

\bibitem[{\citenamefont{Amelino-Camelia}(2013{\natexlab{a}})}]{Amelino-Camelia:2013fxa}
\bibinfo{author}{\bibfnamefont{G.}~\bibnamefont{Amelino-Camelia}},
  \bibinfo{journal}{Phys.Rev.Lett.} \textbf{\bibinfo{volume}{111}},
  \bibinfo{pages}{101301} (\bibinfo{year}{2013}{\natexlab{a}}),
  \eprint{1304.7271}.

\bibitem[{\citenamefont{Majumder}(2013)}]{Majumder:afa}
\bibinfo{author}{\bibfnamefont{B.}~\bibnamefont{Majumder}},
  \bibinfo{journal}{Gen.Rel.Grav.} \textbf{\bibinfo{volume}{11}},
  \bibinfo{pages}{2403} (\bibinfo{year}{2013}), \eprint{1212.6591}.

\bibitem[{\citenamefont{Nozari and Etemadi}(2012)}]{Nozari:2012gd}
\bibinfo{author}{\bibfnamefont{K.}~\bibnamefont{Nozari}} \bibnamefont{and}
  \bibinfo{author}{\bibfnamefont{A.}~\bibnamefont{Etemadi}},
  \bibinfo{journal}{Phys.Rev.} \textbf{\bibinfo{volume}{D85}},
  \bibinfo{pages}{104029} (\bibinfo{year}{2012}), \eprint{1205.0158}.

\bibitem[{\citenamefont{Ching et~al.}(2012)\citenamefont{Ching, Parwani, and
  Singh}}]{Ching:2012vq}
\bibinfo{author}{\bibfnamefont{C.-L.} \bibnamefont{Ching}},
  \bibinfo{author}{\bibfnamefont{R.~R.} \bibnamefont{Parwani}},
  \bibnamefont{and} \bibinfo{author}{\bibfnamefont{K.}~\bibnamefont{Singh}},
  \bibinfo{journal}{Phys.Rev.} \textbf{\bibinfo{volume}{D86}},
  \bibinfo{pages}{084053} (\bibinfo{year}{2012}), \eprint{1204.1642}.

\bibitem[{\citenamefont{Ali}(2014)}]{Ali:2013qza}
\bibinfo{author}{\bibfnamefont{A.~F.} \bibnamefont{Ali}},
  \bibinfo{journal}{Phys.Lett.} \textbf{\bibinfo{volume}{B732}},
  \bibinfo{pages}{335} (\bibinfo{year}{2014}), \eprint{1310.1790}.

\bibitem[{\citenamefont{Ali}(2011)}]{Ali:2011ap}
\bibinfo{author}{\bibfnamefont{A.~F.} \bibnamefont{Ali}},
  \bibinfo{journal}{Class.Quant.Grav.} \textbf{\bibinfo{volume}{28}},
  \bibinfo{pages}{065013} (\bibinfo{year}{2011}), \eprint{1101.4181}.

\bibitem[{\citenamefont{Tawfik et~al.}(2013)\citenamefont{Tawfik, Magdy, and
  Ali}}]{Tawfik:2012he}
\bibinfo{author}{\bibfnamefont{A.}~\bibnamefont{Tawfik}},
  \bibinfo{author}{\bibfnamefont{H.}~\bibnamefont{Magdy}}, \bibnamefont{and}
  \bibinfo{author}{\bibfnamefont{A.~F.} \bibnamefont{Ali}},
  \bibinfo{journal}{Gen.Rel.Grav.} \textbf{\bibinfo{volume}{45}},
  \bibinfo{pages}{1227} (\bibinfo{year}{2013}), \eprint{1208.5655}.

\bibitem[{\citenamefont{Hossenfelder}(2013)}]{Hossenfelder:2012jw}
\bibinfo{author}{\bibfnamefont{S.}~\bibnamefont{Hossenfelder}},
  \bibinfo{journal}{Living Rev.Rel.} \textbf{\bibinfo{volume}{16}},
  \bibinfo{pages}{2} (\bibinfo{year}{2013}), \eprint{1203.6191}.

\bibitem[{\citenamefont{Sprenger et~al.}(2012)\citenamefont{Sprenger, Nicolini,
  and Bleicher}}]{Sprenger:2012uc}
\bibinfo{author}{\bibfnamefont{M.}~\bibnamefont{Sprenger}},
  \bibinfo{author}{\bibfnamefont{P.}~\bibnamefont{Nicolini}}, \bibnamefont{and}
  \bibinfo{author}{\bibfnamefont{M.}~\bibnamefont{Bleicher}},
  \bibinfo{journal}{Eur.J.Phys.} \textbf{\bibinfo{volume}{33}},
  \bibinfo{pages}{853} (\bibinfo{year}{2012}), \eprint{1202.1500}.

\bibitem[{\citenamefont{Amelino-Camelia}(2013{\natexlab{b}})}]{AmelinoCamelia:2008qg}
\bibinfo{author}{\bibfnamefont{G.}~\bibnamefont{Amelino-Camelia}},
  \bibinfo{journal}{Living Rev.Rel.} \textbf{\bibinfo{volume}{16}},
  \bibinfo{pages}{5} (\bibinfo{year}{2013}{\natexlab{b}}), \eprint{0806.0339}.

\bibitem[{\citenamefont{Mukhanov and Chibisov}(1981)}]{mukhanov1981quantum}
\bibinfo{author}{\bibfnamefont{V.~F.} \bibnamefont{Mukhanov}} \bibnamefont{and}
  \bibinfo{author}{\bibfnamefont{G.}~\bibnamefont{Chibisov}},
  \bibinfo{journal}{JETP Letters} \textbf{\bibinfo{volume}{33}},
  \bibinfo{pages}{532} (\bibinfo{year}{1981}).

\bibitem[{\citenamefont{Hawking}(1982)}]{Hawking:1982cz}
\bibinfo{author}{\bibfnamefont{S.~W.} \bibnamefont{Hawking}},
  \bibinfo{journal}{Phys. Lett.} \textbf{\bibinfo{volume}{B115}},
  \bibinfo{pages}{295} (\bibinfo{year}{1982}).

\bibitem[{\citenamefont{Bardeen et~al.}(1983)\citenamefont{Bardeen, Steinhardt,
  and Turner}}]{Bardeen:1983qw}
\bibinfo{author}{\bibfnamefont{J.~M.} \bibnamefont{Bardeen}},
  \bibinfo{author}{\bibfnamefont{P.~J.} \bibnamefont{Steinhardt}},
  \bibnamefont{and} \bibinfo{author}{\bibfnamefont{M.~S.}
  \bibnamefont{Turner}}, \bibinfo{journal}{Phys. Rev.}
  \textbf{\bibinfo{volume}{D28}}, \bibinfo{pages}{679} (\bibinfo{year}{1983}).

\bibitem[{\citenamefont{Guth and Pi}(1982)}]{guth1982fluctuations}
\bibinfo{author}{\bibfnamefont{A.~H.} \bibnamefont{Guth}} \bibnamefont{and}
  \bibinfo{author}{\bibfnamefont{S.-Y.} \bibnamefont{Pi}},
  \bibinfo{journal}{Physical Review Letters} \textbf{\bibinfo{volume}{49}},
  \bibinfo{pages}{1110} (\bibinfo{year}{1982}).

\bibitem[{\citenamefont{Brout et~al.}(1995)\citenamefont{Brout, Massar,
  Parentani, and Spindel}}]{Brout:1995wp}
\bibinfo{author}{\bibfnamefont{R.}~\bibnamefont{Brout}},
  \bibinfo{author}{\bibfnamefont{S.}~\bibnamefont{Massar}},
  \bibinfo{author}{\bibfnamefont{R.}~\bibnamefont{Parentani}},
  \bibnamefont{and} \bibinfo{author}{\bibfnamefont{P.}~\bibnamefont{Spindel}},
  \bibinfo{journal}{Phys. Rev.} \textbf{\bibinfo{volume}{D52}},
  \bibinfo{pages}{4559} (\bibinfo{year}{1995}), \eprint{hep-th/9506121}.

\bibitem[{\citenamefont{Corley and Jacobson}(1996)}]{Corley:1996ar}
\bibinfo{author}{\bibfnamefont{S.}~\bibnamefont{Corley}} \bibnamefont{and}
  \bibinfo{author}{\bibfnamefont{T.}~\bibnamefont{Jacobson}},
  \bibinfo{journal}{Phys. Rev.} \textbf{\bibinfo{volume}{D54}},
  \bibinfo{pages}{1568} (\bibinfo{year}{1996}), \eprint{hep-th/9601073}.

\bibitem[{\citenamefont{Brandenberger and Martin}(2001)}]{Brandenberger:2000wr}
\bibinfo{author}{\bibfnamefont{R.~H.} \bibnamefont{Brandenberger}}
  \bibnamefont{and} \bibinfo{author}{\bibfnamefont{J.}~\bibnamefont{Martin}},
  \bibinfo{journal}{Mod. Phys. Lett.} \textbf{\bibinfo{volume}{A16}},
  \bibinfo{pages}{999} (\bibinfo{year}{2001}), \eprint{astro-ph/0005432}.

\bibitem[{\citenamefont{Kempf}(2001)}]{Kempf:2000ac}
\bibinfo{author}{\bibfnamefont{A.}~\bibnamefont{Kempf}},
  \bibinfo{journal}{Phys.Rev.} \textbf{\bibinfo{volume}{D63}},
  \bibinfo{pages}{083514} (\bibinfo{year}{2001}), \eprint{astro-ph/0009209}.

\bibitem[{\citenamefont{Kempf and Niemeyer}(2001)}]{Kempf:2001fa}
\bibinfo{author}{\bibfnamefont{A.}~\bibnamefont{Kempf}} \bibnamefont{and}
  \bibinfo{author}{\bibfnamefont{J.~C.} \bibnamefont{Niemeyer}},
  \bibinfo{journal}{Phys.Rev.} \textbf{\bibinfo{volume}{D64}},
  \bibinfo{pages}{103501} (\bibinfo{year}{2001}), \eprint{astro-ph/0103225}.

\bibitem[{\citenamefont{Easther et~al.}(2001)\citenamefont{Easther, Greene,
  Kinney, and Shiu}}]{Easther:2001fi}
\bibinfo{author}{\bibfnamefont{R.}~\bibnamefont{Easther}},
  \bibinfo{author}{\bibfnamefont{B.~R.} \bibnamefont{Greene}},
  \bibinfo{author}{\bibfnamefont{W.~H.} \bibnamefont{Kinney}},
  \bibnamefont{and} \bibinfo{author}{\bibfnamefont{G.}~\bibnamefont{Shiu}},
  \bibinfo{journal}{Phys.Rev.} \textbf{\bibinfo{volume}{D64}},
  \bibinfo{pages}{103502} (\bibinfo{year}{2001}), \eprint{hep-th/0104102}.

\bibitem[{\citenamefont{Easther et~al.}(2003)\citenamefont{Easther, Greene,
  Kinney, and Shiu}}]{Easther:2001fz}
\bibinfo{author}{\bibfnamefont{R.}~\bibnamefont{Easther}},
  \bibinfo{author}{\bibfnamefont{B.~R.} \bibnamefont{Greene}},
  \bibinfo{author}{\bibfnamefont{W.~H.} \bibnamefont{Kinney}},
  \bibnamefont{and} \bibinfo{author}{\bibfnamefont{G.}~\bibnamefont{Shiu}},
  \bibinfo{journal}{Phys.Rev.} \textbf{\bibinfo{volume}{D67}},
  \bibinfo{pages}{063508} (\bibinfo{year}{2003}), \eprint{hep-th/0110226}.

\bibitem[{\citenamefont{Ashoorioon et~al.}(2005)\citenamefont{Ashoorioon,
  Kempf, and Mann}}]{Ashoorioon:2004vm}
\bibinfo{author}{\bibfnamefont{A.}~\bibnamefont{Ashoorioon}},
  \bibinfo{author}{\bibfnamefont{A.}~\bibnamefont{Kempf}}, \bibnamefont{and}
  \bibinfo{author}{\bibfnamefont{R.~B.} \bibnamefont{Mann}},
  \bibinfo{journal}{Phys.Rev.} \textbf{\bibinfo{volume}{D71}},
  \bibinfo{pages}{023503} (\bibinfo{year}{2005}), \eprint{astro-ph/0410139}.

\bibitem[{\citenamefont{Ashoorioon and Mann}(2005)}]{Ashoorioon:2004wd}
\bibinfo{author}{\bibfnamefont{A.}~\bibnamefont{Ashoorioon}} \bibnamefont{and}
  \bibinfo{author}{\bibfnamefont{R.~B.} \bibnamefont{Mann}},
  \bibinfo{journal}{Nucl.Phys.} \textbf{\bibinfo{volume}{B716}},
  \bibinfo{pages}{261} (\bibinfo{year}{2005}), \eprint{gr-qc/0411056}.

\bibitem[{\citenamefont{Das and Mann}(2011{\natexlab{b}})}]{Das:2011tq}
\bibinfo{author}{\bibfnamefont{S.}~\bibnamefont{Das}} \bibnamefont{and}
  \bibinfo{author}{\bibfnamefont{R.}~\bibnamefont{Mann}},
  \bibinfo{journal}{Phys.Lett.} \textbf{\bibinfo{volume}{B704}},
  \bibinfo{pages}{596} (\bibinfo{year}{2011}{\natexlab{b}}),
  \eprint{1109.3258}.

\bibitem[{\citenamefont{Lidsey et~al.}(1997)\citenamefont{Lidsey, Liddle, Kolb,
  Copeland, Barreiro, and Abney}}]{Lidsey:1995np}
\bibinfo{author}{\bibfnamefont{J.~E.} \bibnamefont{Lidsey}},
  \bibinfo{author}{\bibfnamefont{A.~R.} \bibnamefont{Liddle}},
  \bibinfo{author}{\bibfnamefont{E.~W.} \bibnamefont{Kolb}},
  \bibinfo{author}{\bibfnamefont{E.~J.} \bibnamefont{Copeland}},
  \bibinfo{author}{\bibfnamefont{T.}~\bibnamefont{Barreiro}}, \bibnamefont{and}
  \bibinfo{author}{\bibfnamefont{M.}~\bibnamefont{Abney}},
  \bibinfo{journal}{Rev. Mod. Phys.} \textbf{\bibinfo{volume}{69}},
  \bibinfo{pages}{373} (\bibinfo{year}{1997}), \eprint{astro-ph/9508078}.

\bibitem[{\citenamefont{Ade et~al.}(2015)}]{Ade:2015lrj}
\bibinfo{author}{\bibfnamefont{P.}~\bibnamefont{Ade}} \bibnamefont{et~al.}
  (\bibinfo{collaboration}{Planck}) (\bibinfo{year}{2015}),
  \eprint{1502.02114}.

\bibitem[{\citenamefont{Bezrukov and Shaposhnikov}(2008)}]{1w}
\bibinfo{author}{\bibfnamefont{F.~L.} \bibnamefont{Bezrukov}} \bibnamefont{and}
  \bibinfo{author}{\bibfnamefont{M.}~\bibnamefont{Shaposhnikov}},
  \bibinfo{journal}{Phys.Lett.} \textbf{\bibinfo{volume}{B659}},
  \bibinfo{pages}{703} (\bibinfo{year}{2008}), \eprint{0710.3755}.

\bibitem[{\citenamefont{Bezrukov et~al.}(2011)\citenamefont{Bezrukov, Magnin,
  Shaposhnikov, and Sibiryakov}}]{w1}
\bibinfo{author}{\bibfnamefont{F.}~\bibnamefont{Bezrukov}},
  \bibinfo{author}{\bibfnamefont{A.}~\bibnamefont{Magnin}},
  \bibinfo{author}{\bibfnamefont{M.}~\bibnamefont{Shaposhnikov}},
  \bibnamefont{and}
  \bibinfo{author}{\bibfnamefont{S.}~\bibnamefont{Sibiryakov}},
  \bibinfo{journal}{JHEP} \textbf{\bibinfo{volume}{1101}}, \bibinfo{pages}{016}
  (\bibinfo{year}{2011}), \eprint{1008.5157}.

\bibitem[{\citenamefont{Kober}(2010)}]{2w}
\bibinfo{author}{\bibfnamefont{M.}~\bibnamefont{Kober}},
  \bibinfo{journal}{Phys.Rev.} \textbf{\bibinfo{volume}{D82}},
  \bibinfo{pages}{085017} (\bibinfo{year}{2010}), \eprint{1008.0154}.

\bibitem[{\citenamefont{Kober}(2011)}]{4w}
\bibinfo{author}{\bibfnamefont{M.}~\bibnamefont{Kober}},
  \bibinfo{journal}{Int.J.Mod.Phys.} \textbf{\bibinfo{volume}{A26}},
  \bibinfo{pages}{4251} (\bibinfo{year}{2011}), \eprint{1104.2319}.

\bibitem[{\citenamefont{Faizal}(2014)}]{5w}
\bibinfo{author}{\bibfnamefont{M.}~\bibnamefont{Faizal}},
  \bibinfo{journal}{Int.J.Geom.Meth.Mod.Phys.} \textbf{\bibinfo{volume}{12}},
  \bibinfo{pages}{1550022} (\bibinfo{year}{2014}), \eprint{1404.5024}.

\bibitem[{\citenamefont{Corless et~al.}(1996)\citenamefont{Corless, Gonnet,
  Hare, Jeffrey, and Knuth}}]{corless1996lambertw}
\bibinfo{author}{\bibfnamefont{R.~M.} \bibnamefont{Corless}},
  \bibinfo{author}{\bibfnamefont{G.~H.} \bibnamefont{Gonnet}},
  \bibinfo{author}{\bibfnamefont{D.~E.} \bibnamefont{Hare}},
  \bibinfo{author}{\bibfnamefont{D.~J.} \bibnamefont{Jeffrey}},
  \bibnamefont{and} \bibinfo{author}{\bibfnamefont{D.~E.} \bibnamefont{Knuth}},
  \bibinfo{journal}{Advances in Computational mathematics}
  \textbf{\bibinfo{volume}{5}}, \bibinfo{pages}{329} (\bibinfo{year}{1996}).

\end{thebibliography}

\end{document}